\documentclass[twocolumn,aps,prl,10pt,superscriptaddress,longbibliography,floatfix,citeautoscript]{revtex4-2}

\usepackage{amsmath}
\usepackage{amssymb}
\usepackage{glossaries}
\usepackage{graphicx}
\usepackage{fancyvrb}
\usepackage{siunitx}
\usepackage[usenames,dvipsnames,svgnames]{xcolor}
\usepackage{hyperref}
\hypersetup{
    pdfnewwindow=true,      
    colorlinks=true,        
    linkcolor=Blue,         
    citecolor=Blue,         
    filecolor=Blue,         
    urlcolor=Blue           
}

\newacronym{dp}{DP}{deep potential}
\newacronym{md}{MD}{molecular dynamics}
\newacronym{mlp}{MLP}{machine-learned potential}
\newacronym{nep}{NEP}{neuroevolution potential}
\newacronym{nqe}{NQE}{nuclear quantum effect}
\newacronym{scan}{SCAN}{strongly constrained and appropriately normed}
\newacronym{pimd}{PIMD}{path-integral molecular dynamics}

\begin{document}

\title{NEP-MB-pol: A unified machine-learned framework for fast and accurate prediction of water's thermodynamic and transport properties}

\author{Ke Xu}
\affiliation{College of Physical Science and Technology, Bohai University, Jinzhou 121013, P. R. China}
\affiliation{Department of Electronic Engineering and Materials Science and Technology Research Center, The Chinese University of Hong Kong, Shatin, N.T., Hong Kong SAR, 999077, P. R. China}

\author{Ting Liang}
\affiliation{Department of Electronic Engineering and Materials Science and Technology Research Center, The Chinese University of Hong Kong, Shatin, N.T., Hong Kong SAR, 999077, P. R. China}

\author{Nan Xu}
\affiliation{College of Chemical and Biological Engineering, Zhejiang University, Hangzhou 310058, P. R. China}

\author{Penghua Ying}
\affiliation{Department of Physical Chemistry, School of Chemistry, Tel Aviv University, Tel Aviv, 6997801, Israel}

\author{Shunda Chen}
\email{phychensd@gmail.com}
\affiliation{Department of Civil and Environmental Engineering, George Washington University,
Washington, DC 20052, USA}

\author{Ning Wei}
\affiliation{Jiangsu Key Laboratory of Advanced Food Manufacturing Equipment and Technology, Jiangnan University, Wuxi, 214122, China}

\author{Jianbin Xu}
\email{jbxu@ee.cuhk.edu.hk}
\affiliation{Department of Electronic Engineering and Materials Science and Technology Research Center, The Chinese University of Hong Kong, Shatin, N.T., Hong Kong SAR, 999077, P. R. China}

\author{Zheyong Fan}
\email{brucenju@gmail.com}
\affiliation{College of Physical Science and Technology, Bohai University, Jinzhou 121013, P. R. China}

\date{\today}

\begin{abstract}
Water's unique hydrogen-bonding network and anomalous properties pose significant challenges for accurately modeling its structural, thermodynamic, and transport behavior across varied conditions. 
Although machine-learned potentials have advanced the prediction of individual properties, a unified computational framework capable of simultaneously capturing water's complex and subtle properties with high accuracy has remained elusive. 
Here, we address this challenge by introducing NEP-MB-pol, a highly accurate and efficient neuroevolution potential (NEP) trained on extensive many-body polarization (MB-pol) reference data approaching coupled-cluster-level accuracy, combined with path-integral molecular dynamics and quantum-correction techniques to incorporate nuclear quantum effects. 
This NEP-MB-pol framework reproduces experimentally measured structural, thermodynamic, and transport properties of water across a broad temperature range, achieving simultaneous, fast, and accurate prediction of self-diffusion coefficient, viscosity, and thermal conductivity.
Our approach provides a unified and robust tool for exploring thermodynamic and transport properties of water under diverse conditions, with significant potential for broader applications across research fields.

\end{abstract}

\maketitle


Water is the foundation of life \cite{lynden_bell_water_2010} and its unique properties make it a central focus of research across numerous scientific disciplines, including physical, chemical, materials, biological, geological, environmental, climate sciences, as well as in energy, food, and technological applications \cite{Levy_annurevbiophys_2006,cremer_NatChem_2018,Ed_NW_2023,Kontogeorgis_review_2022}. Water's structural complexity and anomalous behavior have motivated significant research efforts and extensive studies to characterize and understand its fundamental properties, which are crucial across fields where water serves as a primary or critical component \cite{cowan_nature_2005,russo_NC_2014,sellberg_nature_2014,nilsson_NC_2015,Thamer_science_2015,Shi_PNAS_2020,Kim_science_2020,yu_NC_2020,yang_nature_2021,Flor_Science_2024}.
Despite the critical need to accurately model water's behavior across a wide range of temperatures and pressures, water's intrinsic complexity has made this a long-standing challenge \cite{Pettersson_chemrev_2016}.

Advances in computer simulations of water and aqueous systems have provided powerful tools to study water's properties at the atomistic level, offering insights even beyond experimental capabilities. 
However, a key obstacle in atomistic simulations is achieving both accuracy and computational efficiency. 
Empirical potentials provide low-cost calculations but often lack the fidelity needed to capture water's complex properties, while quantum-mechanical approaches are usually more accurate but come with prohibitive computational costs for large-scale simulations. 
\Glspl{mlp} \cite{behler2007prl,Deringer2019AM,Unke2021cr} have recently transformed this landscape by combining high accuracy with reduced computational cost, allowing simulations that were previously out of reach. Particularly, \glspl{mlp} have shown promise for calculating various properties of water \cite{Omranpour2024jcp}.

Despite this progress, previous works have primarily focused on structural and thermodynamic properties of water \cite{Morawietz2016pnas,Cheng2016JPCL,Singraber2018jpcm,cheng2019pnas,Gartner2020PNAS,Zhang2021PRL,Bore2023NC,chen2024jced}, while relatively few studies address transport properties such as self-diffusion coefficient, viscosity, and thermal conductivity \cite{Morawietz2016pnas, yao2020jcp, malosso2022npj, Tisi2021PRB, Zhang2023jpcb, xu2023accurate}.  
Morawietz \textit{et al.} \cite{Morawietz2016pnas} calculated the self-diffusion coefficient and viscosity of water for a few temperatures using a \gls{mlp} trained using reference data based on revised Perdew-Burke-Ernzerhof (RPBE) or Becke-Lee-Yang-Parr (BLYP) functionals, revealing that van der Waals interactions in the reference data play a crucial role in accurately calculating transport properties. 
However, the results based on both RPBE and BYLP reference data only achieve qualitative agreement with experimental measurements. 
Yao and Kanai \cite{yao2020jcp} calculated the self-diffusion coefficient using a \gls{mlp} trained on \gls{scan} reference data, obtaining results that are significantly lower than experimental values. 
Malosso \textit{et al.} \cite{malosso2022npj} computed viscosity using a \gls{dp} model trained on \gls{scan} reference data, finding that their predictions are significantly overestimated compared to experimental values;  however, a temperature shift could bring the predictions closer to experimental data. 
For thermal transport, both \gls{dp} \cite{Tisi2021PRB,Zhang2023jpcb} and \gls{nep} \cite{xu2023accurate} trained on SCAN reference data have been used to predict the temperature-dependent thermal conductivity. Only the \gls{nep} model, when combined with a quantum correction scheme, achieves quantitative agreement with experimental results \cite{xu2023accurate}.  

The lack of quantitative agreement between calculations and measurements highlights two potential limitations in previous works. 
First, the quality of the reference data has a significant impact on the accuracy of \gls{mlp} predictions, as indeed demonstrated by the markedly different results for transport coefficients \cite{Morawietz2016pnas, yao2020jcp, malosso2022npj} obtained using \glspl{mlp} trained on reference data based on different functionals in density-functional theory calculations.
Second, \glspl{nqe} \cite{Markland2018} play a crucial role in determining water's transport properties. 
While \glspl{nqe} have been shown to impact many static properties of water in atomistic simulations using \glspl{mlp} \cite{Cheng2016JPCL, cheng2019pnas}, the influence of \glspl{nqe} on transport properties remains not fully understood.

\begin{figure*}
\centering
\includegraphics[width=1.6\columnwidth]{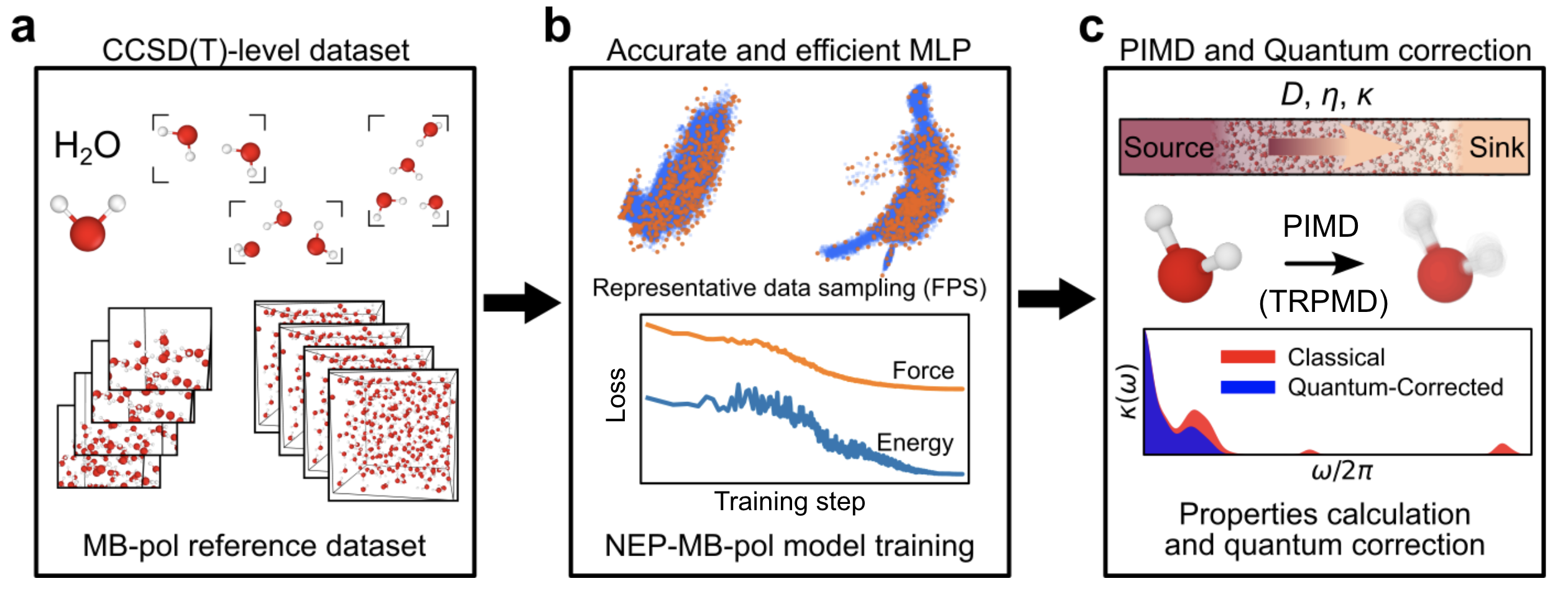}
\caption{\textbf{NEP-MB-pol Workflow/Framework.}
(a) A large dataset was generated using the MB-pol method \cite{Zhai2023JCP, Babin2013JCTC, Babin2014JCTC}, which achieves an accuracy approaching coupled-cluster theory with single, double, and perturbative triple excitations [CCSD(T)]. This dataset includes water monomer, dimer, trimer, tetramer, as well as surface and bulk structures. 
(b) A representative subset of this dataset is selected via farthest-point sampling (FPS) in descriptor space to train a neuroevolution potential (NEP-MB-pol) model, while all the remaining structures are reserved as the test dataset. The NEP-MB-pol model demonstrates low errors for both force and energy on this test dataset.
(c) Utilizing the highly accurate and efficient NEP-MB-pol model, the structural, thermodynamic, and transport properties, including self-diffusion coefficient $D$, viscosity $\eta$, and thermal conductivity $\kappa$, are calculated using molecular dynamics simulations with the versatile graphics processing units molecular dynamics (GPUMD) package. Nuclear quantum effects are accounted for through path-integral molecular dynamics (PIMD) techniques, including thermostatted ring-polymer molecular dynamics (TRPMD), or harmonic quantum correction for spectral thermal conductivity $\kappa(\omega)$.}
\label{fig:fps}
\end{figure*}

\begin{figure*}
\centering
\includegraphics[width=1.6\columnwidth]{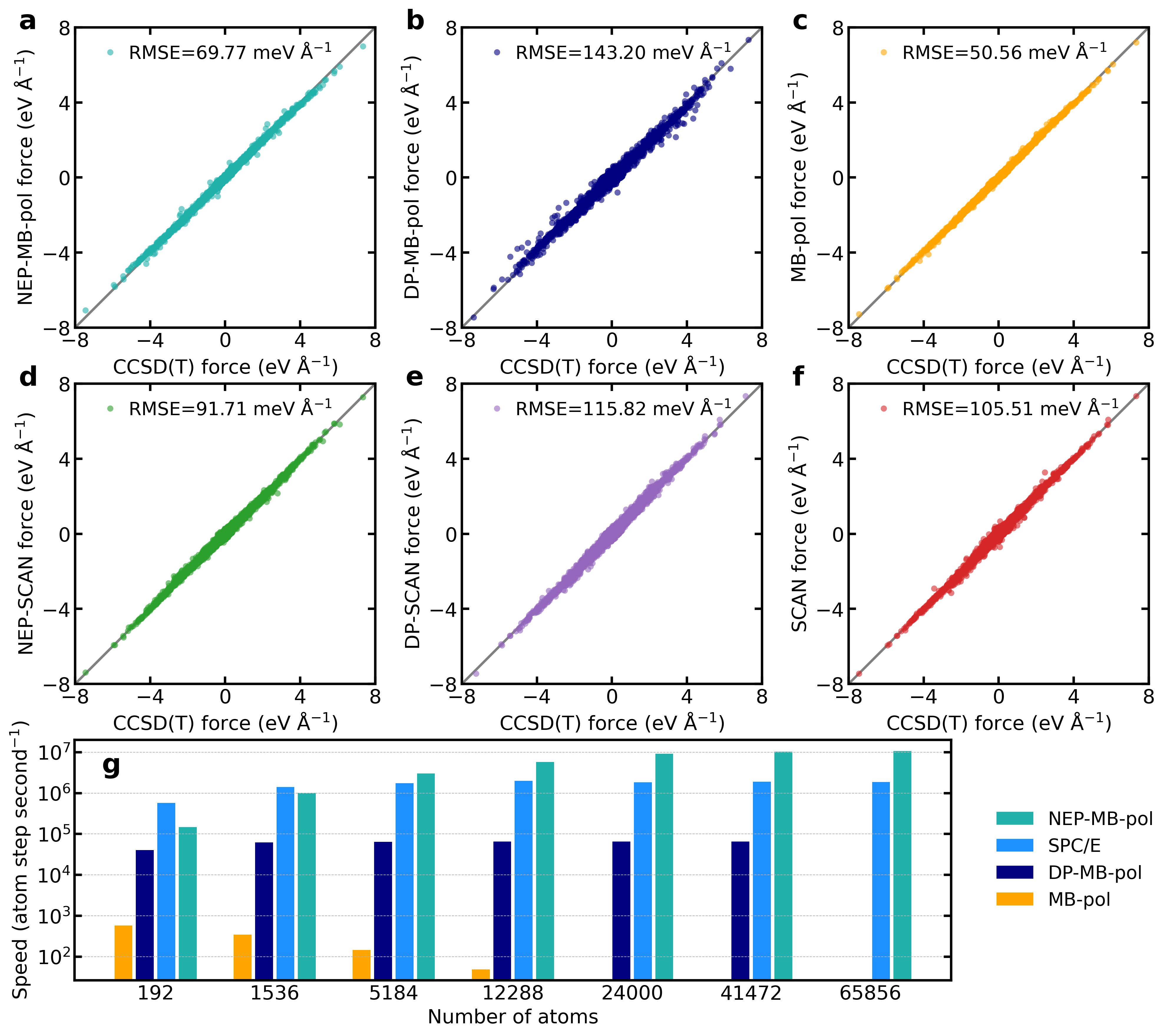}
\caption{\textbf{Evaluation of force accuracy and computational speed.} (a-f) Parity plots comparing reference forces from coupled-cluster theory with single, double, and perturbative triple excitations [CCSD(T)] to forces predicted by various approaches: (a) neuroevolution potential (NEP) model trained on MB-pol dataset (NEP-MB-pol), (b) deep potential (DP) model trained on the MB-pol dataset (DP-MB-pol), (c) MB-pol model, (d) NEP model trained on the the strongly constrained and appropriately normed (SCAN) dataset (NEP-SCAN),  (e) DP model trained on the SCAN dataset (DP-SCAN), and (f) density functional theory (DFT) calculation with the SCAN functional. The root-mean-square error (RMSE) of force for each model is provided. (g) Computational speed as a function of the number of atoms for different water potential models, tested using either 16
i5-14600K CPU cores (for MB-pol and the empirical SPC/E water potential) or a single
RTX 4090 GPU (for NEP-MB-pol and DP-MB-pol).}
\label{fig:com}
\end{figure*}

In this work, we present NEP-MB-pol, a neuroevolution potential model trained on a highly accurate coupled-cluster-level MB-pol dataset, combined with \gls{pimd} \cite{Parrinello1984jcp,ying_NEP-PIMD_2024} and quantum-correction techniques to account for \glspl{nqe}.
MB-pol \cite{Babin2013JCTC, Babin2014JCTC} is built upon the many-body expansion of the interatomic interactions and has been parameterized according to highly accurate quantum-chemistry calculations at the coupled-cluster level, including single, double, and perturbative triple excitations [CCSD(T)], as illustrated in Fig.~\ref{fig:fps}a. 
MB-pol has been demonstrated to be capable of describing the properties of water from the gas to the condensed phases \cite{Medders2014JCTC}.
Although more efficient than typical quantum-mechanical calculations, MB-pol is still too computationally demanding for direct calculation of water's transport properties over a broad temperature range. 
To accelerate these calculations, we employ the \gls{nep} approach \cite{fan2021neuroevolution} to train a \gls{mlp} using MB-pol reference data (Fig.~\ref{fig:fps}b), achieving an interatomic potential model for water with quantum-chemistry-level accuracy and empirical-potential-like speed (Fig.~\ref{fig:com}).
This \gls{nep} model enables extensive large-scale and long-duration \gls{md} simulations that robustly account for \glspl{nqe} (Fig.~\ref{fig:fps}c). For the first time, this NEP-MB-pol framework enables quantitative predictions of all three transport coefficients, including self-diffusion coefficient, viscosity, and thermal conductivity, as well as heat capacity and structural properties like density and radial distribution function across a temperature range simultaneously. 
For comparison, we also train a \gls{nep} model on \gls{scan} reference data, finding it accurately predicts only the thermal conductivity, highlighting the advantage of MB-pol reference data in capturing water's transport behavior.
This NEP-MB-pol framework represents a significant advance in modeling water's thermodynamic and transport properties, with great potential for broader applications.


\textbf{The trained \gls{nep} models.}
The \gls{nep} approach \cite{fan2021neuroevolution} is a framework for generating highly efficient \glspl{mlp}. It is a neural network potential trained using an evolutionary algorithm. 
The site energy $U_i$ of atom $i$ is taken as a function of a descriptor vector with a number of radial and angular descriptors, which are constructed based on the Chebyshev polynomials and the spherical harmonics. 
The descriptors are invariant with respect to translation, rotation, and permutation of atoms of the same species. 
The total energy is obtained by summing the site energies across atoms $U=\sum_i U_i$. The force on an atom is derived as the negative gradient of the total energy with respect to the atom's position, $\mathbf{F}_i = -\partial U/\partial \mathbf{r}_i$. 

\Glspl{mlp} for water based on \gls{dp} have been developed using both MB-pol \cite{Zhai2023JCP} and SCAN \cite{Zhang2021PRL} reference data. Both datasets are very comprehensive. 
The MB-pol dataset \cite{Zhai2023JCP} includes \num{59181} molecular structures (water monomers, dimers, etc), \num{43494} bulk water structures, and 1932 bulk-molecule interface structures, totaling \num{104607} structures (Fig.~\ref{fig:fps}a). 
The SCAN dataset \cite{Zhang2021PRL} contains \num{48419} structures sampled from \gls{md} simulations with a wide range of temperature (150 to 2000 K) and pressure (10$^{-4}$ to 50 GPa). 
In this work, we utilize these two datasets, with sub-sampling as detailed below, to train \gls{nep} models.

In principle, we could train \gls{nep} models using the full datasets that were originally used for training the \gls{dp} models \cite{Zhang2021PRL,Zhai2023JCP}. 
However, this is not an optimal strategy for the very data-efficient \gls{nep} approach, which is based on smaller-scale neural network models. 
By employing farthest-point sampling in the descriptor space, we obtained significantly reduced yet representative training datasets, containing only 1\% of the original full datasets (1250 and 601 structures for the reduced MB-pol and SCAN datasets, respectively). 
These reduced datasets are visualized via principal component analysis (PCA) of descriptor space, as shown in Fig.~\ref{fig:fps}b and Fig. S1.
The \gls{nep} models trained on these reduced datasets are referred to as NEP-MB-pol and NEP-SCAN, respectively, while the \gls{dp} models trained on the full datasets are denoted as DP-MB-pol \cite{Zhai2023JCP} and DP-SCAN \cite{Zhang2021PRL}, respectively.
One advantage of using a reduced training dataset is availability of many unseen structures, which enables robust validation against over-fitting.
For NEP-MB-pol, the root-mean-square errors of energy and force evaluated on the entire dataset are 1.2 meV atom$^{-1}$ and 45.4 meV \AA$^{-1}$, respectively. 
In comparison, the corresponding values for \gls{dp}-MB-pol \cite{Zhai2023JCP} (specifically, the model labeled DNN-seed2) are 23.3 meV atom$^{-1}$ and 48.2 meV \AA$^{-1}$, respectively (see Fig. S2 for parity plots).
For NEP-SCAN, the root-mean-square errors of energy and force evaluated on the whole dataset are 3.3 meV atom$^{-1}$ and 85.1 meV \AA$^{-1}$, respectively, while the corresponding values for \gls{dp}-SCAN \cite{Zhang2021PRL} are 2.3 meV atom$^{-1}$ and 121.1 meV \AA$^{-1}$, respectively (see Fig. S3 for parity plots). 
These comparisons demonstrate that the \gls{nep} models make highly accurate predictions for unseen structures.
Another benefit of using a reduced training dataset is flexibility for future extensions; adding a small number of structures into a large existing dataset typically does not lead to significant improvements, while a reduced dataset allows for more effective extension.

\textbf{Further accuracy validation of NEP models.}
To further evaluate the accuracy of our \gls{nep} models, we generated an independent validation dataset at the CCSD(T) level of theory (see Methods for details). Using CCSD(T) as the reference gold standard, we present parity plots (Fig.~\ref{fig:com}a-f) that compare reference forces from CCSD(T) with those predicted by various approaches. 
Our results demonstrate that MB-pol is indeed significantly more accurate than density-functional theory calculations using the SCAN functional (see Methods for details).
Consequently, the NEP-MB-pol model, trained on the MB-pol dataset, is more accurate than the NEP-SCAN model, which was trained on the SCAN dataset. Among the four \glspl{mlp}, our NEP-MB-pol model stands out, achieving the best accuracy, with a force root-mean-square error of 69.77 meV \AA$^{-1}$, which is only slightly higher than the MB-pol model's value of 50.56 meV \AA$^{-1}$. For comparison, the DP-MB-pol \cite{Zhai2023JCP} and DP-SCAN \cite{Zhang2021PRL} models are found to be less accurate.

\textbf{Computational speed of NEP-MB-pol.}
The computational efficiency of a \gls{mlp} is crucial for its effective applications in large-scale and long-duration \gls{md} simulations, especially when numerous replicas per atom are needed to capture the \glspl{nqe}. 
The \gls{nep} model, as implemented in \textsc{gpumd} \cite{fan2017cpc}, exhibits excellent computational performance.
Our \gls{nep} model achieves a computational speed of about \num{1e7} atom step s$^{-1}$ in \gls{md} simulations with systems containing more than \num{12000} atoms, using a single GeForce RTX 4090 GPU card (Fig.~\ref{fig:com}g). 
This is about 100 times faster than \gls{dp} running on the same GPU and is a few times faster than the SPC/E empirical potential \cite{Berendsen1987JPC} running on 16 i5-14600K CPU cores. 
It is important to note that the MB-pol model can only simulate up to about \num{12000} atoms and its computational cost scales quadratically with respect to system size.
At this scale, \gls{nep}-MB-pol is about 5 orders of magnitude faster than MB-pol.
This dramatic speedup empowers us with a water potential model with quantum-chemical accuracy and empirical-potential-like speed.

\vspace{0.5cm}

\begin{figure*}
\centering
\includegraphics[width=1.8\columnwidth]{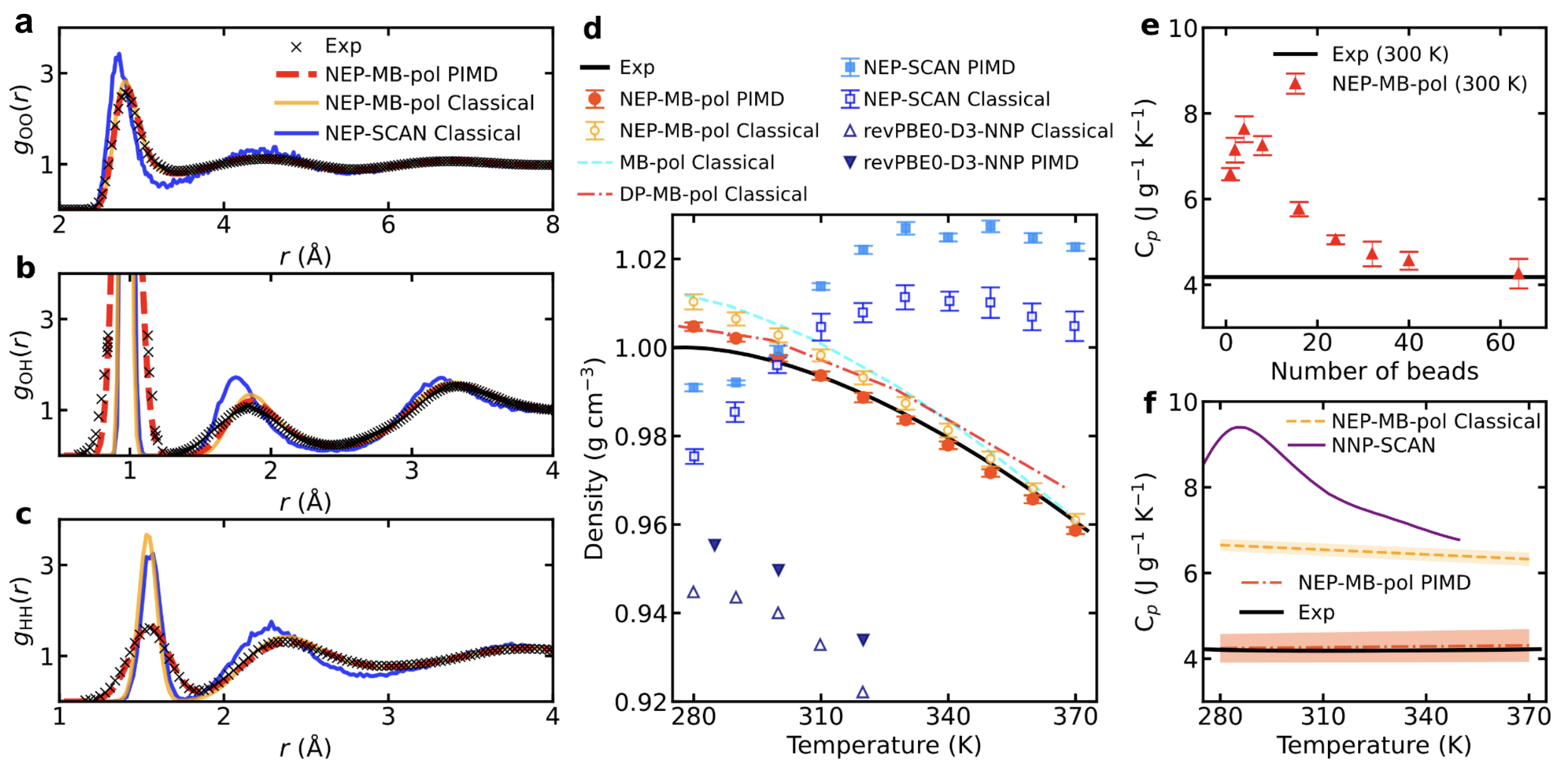}
\caption{
\textbf{Radial distribution functions, density, and isobaric heat capacity of water.} Radial distribution functions of (a) O-O atom pairs, (b) O-H atom pairs and (c) H-H atom pairs. Results from path-integral molecular dynamics (PIMD) simulations with NEP-MB-pol using 32 beads at 300 K (NEP-MB-pol PIMD, red dashed line) agree well with experimental data for O-O (295.1 K) \cite{Skinner2013JCP}, O-H (300 K) \cite{SoperCP2000} and H-H (300 K) \cite{SoperCP2000} atom pairs. For comparison, results from classical molecular dynamics (MD) simulations with NEP-MP-pol (NEP-MP-pol Classical, orange line) and NEP-SCAN (NEP-SCAN Classical, blue line) are also provided.
(d) Density of water as a function of temperature, comparing experimental data (solid line) \cite{Nist_chemistry_webbook, Wagner2002JPCRD}, classical MD simulations with the MB-pol model \cite{Riera2023JCP,Paesani2016ACR} (dashed line), the DP-MB-pol model \cite{Zhai2023JCP} (dot-dashed line), the revPBE0-D3-NNP \cite{cheng2019pnas} (open triangles), the NEP-MB-pol model (open circles), and the NEP-SCAN model (open squares), as well as PIMD simulations with NEP-MB-pol model using 32 beads (filled circles), NEP-SCAN model (filled squares), and the revPBE0-D3-NNP \cite{cheng2019pnas} (inverted filled triangles; error bars for NEP-MB-pol and NEP-SCAN are based on the standard error of 50 ps data sampled from MD simulations. 
(e) Isobaric heat capacity C$_p$ of liquid water at 300 K as a function of the number of beads, calculated with the NEP-MB-pol model, showing good convergence at 64 beads. (f) C$_p$ as a function of temperature, comparing experimental data \cite{Nist_chemistry_webbook, Wagner2002JPCRD} (solid black line), classical MD simulations with NNP-SCAN \cite{Piaggi2021JCTC} (purple line) and  NEP-MB-pol (NEP-MB-pol Classical, orange dashed line), as well as PIMD simulations with the NEP-MB-pol model using 64 beads (NEP-MB-pol PIMD, red dotted-dashed line). The shaded error band around the line represents uncertainty.
}
\label{fig:rdfdencp}
\end{figure*}

\vspace{0.5cm}

\noindent{\textbf{Structural and thermodynamic properties.}} 
Prominent features of water can be reflected by the radial distribution function $g(r)$. For O-O pairs, NEP-MB-pol can very accurately reproduce the experimental data, even at the classical \gls{md} level, with \gls{pimd} introducing minimal changes. This indicates that \glspl{nqe} are negligible for oxygen. In contrast, NEP-SCAN overshoots both the first and the second peaks of the O-O distribution. However, the O-H and H-H distributions exhibit very strong \glspl{nqe}. With classical \gls{md}, both NEP-MB-pol and NEP-SCAN significantly underestimate the width of the first peaks in $g_{\rm OH}$ and $g_{\rm HH}$, reflecting the absence of zero-point motion in classical \gls{md} simulations. 
Remarkably, \gls{pimd} brings NEP-MP-pol results much closer to experimental data, particularly for $g_{\rm HH}$. 
Note that the small differences around the second peak in $g_{\rm OH}$ from NEP-MB-pol with classical \gls{md} have also been observed in the prior work using MB-pol \cite{Medders2014JCTC}. On the other hand, the results from the NEP-SCAN model with classical \gls{md} show more pronounced deviations from the experiential data, which cannot be totally attributed to the absence of \glspl{nqe}. The excellent agreement between NEP-MB-pol with \gls{pimd} and experimental data highlights both the high accuracy of our NEP-MB-pol model and the critical role of \glspl{nqe} in describing O-H and H-H bonds. 

The accurate prediction of density further demonstrates the reliability of the NEP-MB-pol potential. 
In the temperature range from 280 to 370 K at 1 atm pressure, the density of water monotonically decreases according to experimental data \cite{Nist_chemistry_webbook, Huber2009JPCRD}, which have very small uncertainties (Fig.~\ref{fig:rdfdencp}d). 
NEP-MB-pol, even with classical \gls{md} simulations, reproduces this trend well.
The agreement between NEP-MB-pol and experiments is further improved by considering \glspl{nqe} using \gls{pimd} simulations. 
This is consistent with the radial distribution function results: \glspl{nqe} result in lower density at a given temperature. 
The maximum difference between \gls{pimd} simulation with NEP-MB-pol and experiments is at 280 K, which is about $0.4\%$.
We have checked that this difference cannot be made smaller by using a larger number of replicas in the \gls{pimd} simulations (Fig. S4).
It thus represents a small degree of inaccuracy of the NEP-MB-pol model, likely inherited from the MB-pol model. Indeed, density predictions using the original MB-pol model \cite{Riera2023JCP, Paesani2016ACR} at the classical \gls{md} level agree well with our NEP-MB-pol results. 
Notably, the classical MD results from DP-MB-pol \cite{Zhai2023JCP} show poor agreement with those from MB-pol, reflecting the relatively higher accuracy of our NEP-MB-pol model over the DP-MB-pol model.

Figure~\ref{fig:rdfdencp}d also presents density results obtained by classical \gls{md} and \gls{pimd} simulations using the NEP-SCAN model. 
It is evident that the NEP-SCAN model predicts a drastically different trend than experiments, significantly overestimating the density when $T> 300$ K. 
This is consistent with the overestimation of the melting point by \gls{scan}, which originates from its overestimated strength of hydrogen bonds \cite{Piaggi2021JCTC}.

Figure~\ref{fig:rdfdencp}e shows the heat capacity $C_p$ of water as a function of the number of beads, as predicted by the NEP-MB-pol model. For systems with 64 beads, $C_p$ obtained from NEP-MB-pol is in excellent agreement with experimental data \cite{Nist_chemistry_webbook, Huber2009JPCRD}. The relationship between $C_p$ and temperature is presented in Fig.~\ref{fig:rdfdencp}f. As observed, predictions from \gls{pimd} simulations with NEP-MB-pol, incorporating \glspl{nqe}, match experimental observations very well. In contrast, predictions from classical MD simulations with NEP-MB-pol show noticeable deviations, and the results from classical MD simulations with NNP-SCAN \cite{Piaggi2021JCTC} exhibit even greater deviations. These results highlight the importance of using a high-quality dataset and accurately incorporating \glspl{nqe} to achieve precise predictions of water's heat capacity.

\vspace{0.5cm}

\begin{figure*}
\centering
\includegraphics[width=2\columnwidth]{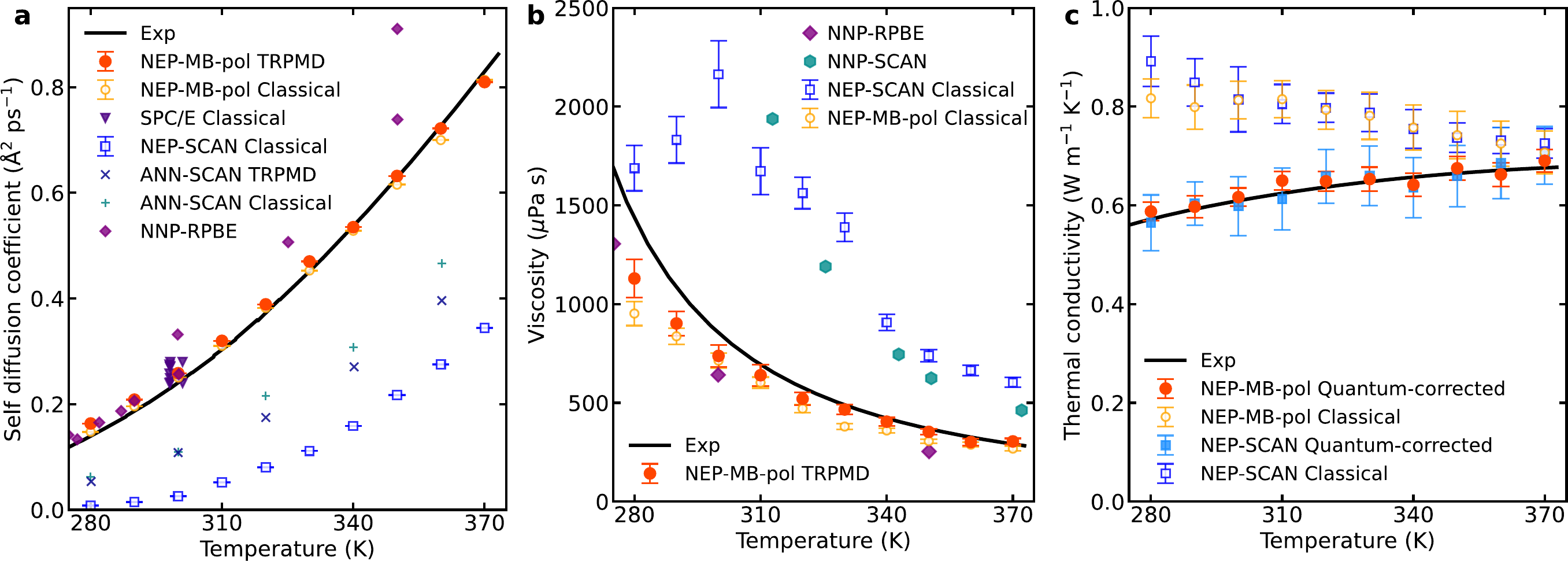}
\caption{\textbf{Transport coefficients of water.}
    (a) Self diffusion coefficient of water as a function of temperature from experiments (solid line) \cite{Holz2000PCCP}, classical molecular dynamics (MD) simulations with the SPC/E model \cite{Guevara2011JCP} (filled triangles), the NEP-SCAN model (open squares), the NEP-MB-pol model (open circles), the ANN-SCAN model \cite{yao2020jcp} (plus symbols), the NNP-RPBE model \cite{Morawietz2016pnas,de_hijes_2024JCP,Montero2023JCP} 
 (filled diamond), and thermostatted ring-polymer molecular dynamics (TRPMD) simulations with the NEP-MB-pol model using 32 beads (filled circles) and the ANN-SCAN model \cite{yao2020jcp} (cross symbols). The error bars in NEP-MB-pol and NEP-SCAN results are calculated from the standard error of the mean for 10 independent simulations.
    (b) Viscosity of water as a function of temperature from experiments (solid line) \cite{Nist_chemistry_webbook, Huber2009JPCRD}, classical MD simulations with the NEP-SCAN model (open squares), the NEP-MB-pol model (open circles), NNP-RPBE model \cite{Morawietz2016pnas} (filled diamond), NNP-SCAN \cite{malosso2022npj} (filled hexagon), and TRPMD simulations with the NEP-MB-pol model using 32 beads (filled circles). The error bars in NEP-MB-pol and NEP-SCAN results are based on the standard error of the mean for 30 independent simulations.
    (c) Thermal conductivity of water as a function of temperature from experiments (solid line) \cite{Nist_chemistry_webbook, Huber2012JPCRD}, classical MD simulations with the NEP-SCAN model (open squares) and the NEP-MB-pol model (open circles). The filled circles and squares represent the NEP-MB-pol and NEP-SCAN results after using the quasi-harmonic quantum correction. The error bars are calculated as standard errors from three independent homogeneous nonequilibrium MD simulations, each with 3 ns production time, and 100 equilibrium MD simulations, each with 10 ps production time.
    }
\label{fig:sdcvisctc}
\end{figure*}

\noindent{\textbf{Self-diffusion coefficient.}}
We now move on to the study of water's transport properties, starting from the self-diffusion coefficient. 
The diffusion coefficient can be calculated either as the time derivative of the mean-square displacement, or equivalently, the time integral of the velocity autocorrelation function. 
The running diffusion coefficient as a function of the correlation time calculated using the NEP-MB-pol model in the temperature range of 280 to 370 K (all at 1 atm) are shown in Fig. S5. 
The running diffusion coefficient converge well up to a correlation time of 5 ps, which is then taken as the upper limit in the time integral.
Finite-size effects may influence diffusion coefficient calculations in \gls{md} simulations. 
We have carefully checked the finite-size effects in Fig. S7, and found that the finite-size effects are negligible when the number of atoms in the simulation cell exceeds \num{10000}. 
We have chosen a very safe calculation cell with \num{24567} atoms in all the diffusion coefficient and subsequent transport calculations to ensure accuracy. 

The time-converged diffusion coefficient values from 280 to 370 K are presented in Fig.~\ref{fig:sdcvisctc}a. 
Using classical \gls{md} simulations, NEP-MB-pol already achieves a good agreement with experimental results \cite{Easteal1989JCS, mills1973JPC, Holz2000PCCP}.
Incorporating \glspl{nqe} using thermostatted ring-polymer \gls{md} \cite{rossi_how_2014,ying_NEP-PIMD_2024} further improves the agreement, particularly at higher temperatures. 
Minor deviations between NEP-MB-pol results and experiment data likely stem from the small inaccuracies in the NEP-MB-pol model, as noted earlier.

Previous theoretical calculations using empirical potentials, such as the SPC/E model \cite{Guevara2011JCP, Ioannis2019TF}, have typically focused on narrower temperature ranges near room temperature. 
While these empirical potentials may perform well at specific temperatures, they often struggle to match experimental results across a wide range of temperatures. 

Prior theoretical predictions of the diffusion coefficient of water using \glspl{mlp} also exhibit limitations. For example, using classical \gls{md} simulations, the predicted diffusion coefficients from the ANN potential trained with \gls{scan} reference data (ANN-SCAN) \cite{yao2020jcp} are significantly underestimated across the whole temperature range. 
This trend is also confirmed by our NEP-SCAN results.
The quantum effects as revealed by the ANN-SCAN simulations are also rather small.
Using another NNP trained using RPBE (NNP-RPBE) reference data \cite{Morawietz2016pnas}, the calculated diffusion coefficient values from classical \gls{md} simulations significantly overestimate the experimental results. 
Given the minimal overall \glspl{nqe} in diffusion coefficient calculations, the results in Fig.~\ref{fig:sdcvisctc}a strongly suggest that NEP-MB-pol has a higher accuracy than previous \glspl{mlp}.

The capability of accurately predicting the diffusion coefficient across a wide range of temperatures using NEP-MB-pol suggests that this model has prediction power and our computational framework could be very useful for exploring water properties at extreme conditions that are inaccessible to experimental measurements. 

\vspace{0.5cm}

\noindent{\textbf{Shear viscosity.}}
We next examine shear viscosity $\eta$ of water, which can be calculated as a time integration of the shear stress autocorrelation function (see Method for details). 
We consider the temperatures from 280 to 370 K and a constant pressure of 1 atm. The running shear viscosity as a function of the correlation time calculated using the NEP-MB-pol model for selected temperatures are shown in Fig. S6. 
Similar to the running diffusion coefficient, the running viscosity $\eta(t)$ also converges with respect to the correlation time $t$. 
One difference between diffusion coefficient and viscosity is that the latter involves a collective correlation function, which exhibits higher statistical error with the same length of trajectory. 
Therefore, we need to perform more independent runs for viscosity calculations. Here we have performed 30 independent runs for each temperature. 

The time-converged shear viscosity values calculated from NEP-MB-pol and NEP-SCAN are presented in Fig.~\ref{fig:sdcvisctc}b. The results are compared with experimental data \cite{Nist_chemistry_webbook,Huber2009JPCRD} and previous results from DP-SCAN \cite{malosso2022npj}.
Using classical \gls{md} simulations, the NEP-MB-pol model produces results generally smaller than experimental values. 
The maximum deviation is about 40\% at 280 K.
Incorporating \glspl{nqe} via thermostatted ring-polymer \gls{md} simulations significantly improves the calculated results, although there is still an underestimation of about 20\% at 280 K. 
This underestimation is likely related to the small inaccuracy of the NEP-MB-pol mentioned above. 
This means that viscosity is a physical property that is highly sensitive to the accuracy of the interatomic force in the \gls{md} simulations. 
Indeed, the NNP-RPBE model with classical \gls{md} simulations \cite{Morawietz2016pnas} shows a larger underestimation than our MB-NEP-pol model, while both the NEP-SCAN model and the previous DP-SCAN model predict viscosity values that are several times larger than the experimental results.

\vspace{0.5cm}

\noindent{\textbf{Thermal conductivity.}}
Finally, we investigate the thermal conductivity $\kappa$ of water, which can be calculated as a time integral of the heat current auto-correlation function. Because the heat current can be decomposed into potential (p) and kinetic (k) parts, the heat current auto-correlation function, and hence the running thermal conductivity, can be decomposed into three terms: the p-p term ($\kappa^{pp}$), the k-k term ($\kappa^{kk}$), and the cross term ($\kappa^{pk}$). The running thermal conductivity for the three terms all converges well up to a correlation time of 1 ps, as shown in Fig. S8. We observe that the cross term $\kappa^{pk}$ is essentially zero and the k-k term $\kappa^{kk}$ contributes a small but non-negligible portion.

The total thermal conductivity from both NEP-MB-pol and NEP-SCAN calculated from 280 to 370 K (with a constant pressure of 1 atm) are shown in Fig.~\ref{fig:sdcvisctc}c. It is clear that both sets of results are quite consistent with each other and are significantly overestimated compared to the experimental results. The deviation between calculations and experiments increases at lower temperatures, which again indicates strong \glspl{nqe}. Unfortunately, the heat current is a nonlinear operator and there are so far no feasible path-integral techniques that can account for the \glspl{nqe} in thermal conductivity calculations. 
However, we notice that the simpler quantum-correction method based on harmonic approximation \cite{Berens1983JCP} has been proven to be a feasible one for disordered materials, which naturally include liquid water \cite{xu2023accurate}. Here, the harmonic approximation means that one assumes that \glspl{nqe} are negligible for the anharmonic characteristics in a system. 
This can be well confirmed by the calculation of heat capacity \cite{Berens1983JCP}. 
As a prerequisite for applying this quantum-correction scheme, we need to first obtain a spectral decomposition $\kappa(\omega)$ of the thermal conductivity, which, fortunately, can be conveniently achieved within the homogeneous nonequilibrium \gls{md} formalism \cite{fan2019homogeneous}. 
Here the part that needs to be quantum-corrected is the p-p component, which becomes $\kappa^{pp}(\omega)x^2e^x/(e^x-1)^2$ after quantum correction, where $x=\hbar \omega / k_{\rm{B}} T$ and $\hbar$ is the reduced Planck constant.
Adding up the quantum corrected $\kappa^{pp}$ and the original $\kappa^{kk}$ gives the total thermal conductivity that agrees well with experiments as shown in Fig.~\ref{fig:sdcvisctc}c. 

\vspace{0.5cm}

\textbf{Discussion.}
In this work, we achieved quantitative predictions of a broad range of physical properties of liquid water across various temperatures.
The physical properties evaluated include not only structural characteristics, such as density and radial distribution functions, but also thermodynamic properties such as isobaric heat capacity, and transport properties including the self diffusion coefficient, viscosity, and thermal conductivity.
The high level of agreement between our calculations and experiments can be attributed to two primary factors. 

First, we developed the highly accurate and efficient NEP-MB-pol model, which serves as the foundation of this work.
The high accuracy of our NEP-MB-pol model, further validated by using an independent CCSD(T) dataset prepared in this study, is inherited from the underlying training data based on the MB-pol model, which has been shown to be highly accurate \cite{Riera2023JCP,Paesani2016ACR}. 
While the MB-pol model itself is very computationally intensive and has been impractical for calculating transport properties that require extensive simulations, our NEP-MB-pol model not only retains MB-pol's high accuracy but also achieves computational speeds several orders of magnitude faster, making extensive large-scale simulations feasible. 

Second, the reliable prediction of water's properties benefits from the use of appropriate methods that accurately account for nuclear quantum effects, which are especially strong in water.
For static properties like density (equation of state), path-integral molecular dynamics correctly captures the nuclear quantum effects related to zero-point motion, resulting in density values that align closely with experimental observations.
For dynamic properties, thermostatted ring-polymer molecular dynamics effectively captures diffusion and viscosity calculations, while the quasi-harmonic quantum corrections address overpopulated high-frequency vibrations in the calculation of thermal conductivity. 

An important aspect of our approach is that our NEP-MB-pol model was developed without fitting to experimental data, yet it accurately predicts multiple experimentally validated properties. 
Our approach distinguishes itself from empirical models, which are often tailored  to fit specific experimental properties and may have limited applicability across different thermodynamic conditions.
By relying on first principles calculations and machine learning, the predictive power of our NEP-MB-pol model is constrained only by the accuracy of the MB-pol model and the breadth and diversity of the training data.

To further improve the model's accuracy, additional training data from direct CCSD(T) calculations could be incorporated. The NEP-MB-pol framework is also compatible with other coupled-cluster-level datasets from, for example, MB-pol(2023) \cite{MB-pol2023_jctc_2023,Palos_jctc_MB-pol_review_2024} and q-AQUA \cite{q-AQUA_JPCL_2022} models, which include higher-order interactions. Additionally, the NEP-MB-pol framework is adaptable to a variety of machine learning potentials, for enhancing their predictive capabilities in modeling water.

While this study focused on the water properties under ambient conditions, our approach is extendable. With an extended training dataset, NEP-MB-pol has the potential to model water's behavior under more extreme thermodynamic conditions.
In conclusion, we believe that NEP-MB-pol represents a versatile and scalable approach with promising applications for exploring the unique properties and phenomena of water and related systems across multiple fields.

\section{Methods}

\noindent{\textbf{The NEP model.}}
In the NEP approach, the site energy $U_{i}$ of atom $i$ can be written as
\begin{equation}
\label{equation:Ui}
U_i = \sum_{\mu=1}^{N_\mathrm{neu}}w^{(1)}_{\mu}\tanh\left(\sum_{\nu=1}^{N_\mathrm{des}} w^{(0)}_{\mu\nu} q^i_{\nu} - b^{(0)}_{\mu}\right) - b^{(1)},
\end{equation}
where $\tanh(x)$ is the activation function, $\mathbf{w}^{(0)}$, $\mathbf{w}^{(1)}$, $\mathbf{b}^{(0)}$, and $b^{(1)}$ are the weight and bias parameters. 
The descriptor $q_i^{\nu}$ is an abstract vector whose components group into radial and angular parts.
The radial descriptor components $q_n^i $ $(0 \leq {n} \leq n_{\rm max}^{\rm R})$ are defined as
\begin{equation}
\label{equation:rad_des}
q_{n}^i = \sum_{j\neq{i}}g_n(r_{ij}), 
\end{equation}
where $r_{ij}$ is the distance between atoms $i$ and $j$ and $g_n(r_{ij})$ are a set of radial functions, each of which is formed by a linear combination of Chebyshev polynomials. 
The angular components include $n$-body ($n= 3,4,5$) correlations. 
For the 3-body part, the descriptor components are defined as $(0 \leq {n} \leq n_{\rm max}^{\rm A}$,  $1 \leq {l} \leq l_{\rm max}^{\rm 3body})$ 
\begin{equation}
\label{equation:ang_des}
q_{nl}^i = \sum_m (-1)^m A_{nlm}^i A_{nl(-m)}^i;
\end{equation}
\begin{equation}
A_{nlm}^i = \sum_{j\neq i} g_n(r_{ij}) Y_{lm}(\hat{\mathbf{r}}_{ij}).
\end{equation}
Here, $Y_{lm}$ are the spherical harmonics and $\hat{\mathbf{r}}_{ij}$ is the unit vector of $\mathbf{r}_{ij}$. Note that the radial functions $g_n(r_{ij})$ for the radial and angular descriptor components can have different cutoff radii, which are denoted as $r_{\rm c}^{\rm R}$ and $r_{\rm c}^{\rm A}$, respectively. 

\noindent{\textbf{Generating a validation dataset at CCSD(T) level of theory.}
The CFOUR program package \cite{Matthews_JCP_2020_CFOUR}, with aug-cc-pVTZ (aVTZ) basis set, was utilized to generate an independent validation dataset at the coupled-cluster level of theory, including single, double, and perturbative triple excitations [CCSD(T)]. 
This validation dataset comprises a total of 262 structures, including 56 structures (each containing up to six water molecules) selected from various publications \cite{CM2024ET,Temelso2011JPCA,Malloum2019NJC}, and 206 structures sampled by \gls{md} simulations driven by the NEP-MB-pol model.

\noindent{\textbf{Density-functional theory calculations using SCAN functional.}} 
Density-functional theory calculations using the strongly constrained and appropriately normed (SCAN) functional were performed with the Vienna Ab initio Simulation Package (VASP, version 6.3.0) \cite{Kresse1996PRB}, to predict the energies, forces, and virial of the 262 structures in the independent validation CCSD(T) dataset (see above). To account for the non-spherical contributions to the gradient correction within the projector-augmented-wave sphere, the flag LASPH was set to TRUE. A kinetic energy cutoff of 1500 eV was applied for the plane waves, and a reciprocal space sampling grid spacing of 0.5 \AA$^{-1}$ was used. The self-consistent field convergence threshold was set to 10$^{-6}$ eV.

\noindent{\textbf{Molecular dynamics simulations.}} 
For all the MD simulations conducted to compute the physical properties reported in this study, the simulation system consists of \num{24567} atoms in a periodic cubic box with dimensions of about 6.2 nm in each direction. The system pressure is maintained at 1 atm.
The time step of 0.5 fs is used for integration in classical, path-integral, and thermostatted ring-polymer molecular dynamics simulations. 
To calculate the density of water at each temperature, the system is equilibrated for 50 ps in the isothermal-isobaric ensemble.
The pressure of the system is still kept at 1 atm, while the temperature is varied from 280 K to 370 K in increments of 10 K. 
All the molecular dynamics simulations were performed using the graphics processing units molecular dynamics (GPUMD) package \cite{fan2017cpc}.

\noindent{\textbf{Isobaric heat capacity.}}
The isobaric heat capacity C$_p$ is defined as the rate of change of enthalpy $H$ with respect to temperature at constant pressure:
\begin{equation}
\label{equ:cp1}
C_p = \left(\frac{\partial H}{\partial T}\right)_p
\end{equation}
where, $H=E+pV$, and $E$, $T$, $p$, and $V$ represent the internal energy, temperature, pressure, and volume, respectively. 
To compute $C_p$, the value of $H$ is calculated at a series of temperature points with the same pressure $p$. A quadratic function is then fitted to the relationship between $H$ and $T$, and its first derivative yields $C_p$ as a function of $T$.

\noindent{\textbf{Self-diffusion coefficient.}} The running self-diffusion coefficient for water is calculated using the following Green-Kubo relation:
\begin{equation}
\label{equ:vacf}
D(t) = \frac{1}{3} \int_{\tau=0}^{t} C_{vv}(\tau) \rm{d} \tau
\end{equation}
where the velocity auto-correlation function is defined as
$
    C_{vv}(\tau) = \frac{1}{N} \sum_i^N \langle \mathbf{v}_i(0) \cdot \mathbf{v}_i(\tau) \rangle
$.
Here, $N$ is the number of atoms in the systems and $\mathbf{v}_i$ is the velocity of atom $i$.

\noindent{\textbf{Shear viscosity.}}
The shear viscosity is defined as 
$
\eta = \frac{1}{3} \left( \eta_{xy} + \eta_{xz} + \eta_{yz} \right)
$,
where the running integral of $\eta_{\alpha\beta}$ is calculated using the following Green-Kubo relation:
\begin{equation}
\eta_{\alpha \beta}(t) = \frac{V}{k_{\rm B}T}\int_0^t C_{pp}(\tau) \rm{d} \tau.
\end{equation}
Here, 
$
C_{pp}(\tau) = \langle (p_{\alpha \beta} (0)- \langle p_{\alpha \beta}\rangle) (p_{\alpha \beta} (\tau) - \langle p_{\alpha \beta}\rangle) \rangle
$
is the pressure auto-correlation function, 
$V$ is the volume, $k_{\rm B}$ is Boltzmann's constant, $T$ is temperature, and $p_{\alpha\beta}$ is the pressure tensor.

\noindent{\textbf{Thermal conductivity.}}
Similarly, we can use a Green-Kubo relation to calculate thermal conductivity:
\begin{equation}
\label{equation:gk}
\kappa(t)=\frac{1}{k_{\rm B}T^2V}\int_0^t \langle \mathbf{J}(0) \cdot \mathbf{J}(\tau)\rangle d\tau,
\end{equation}
where $\mathbf{J}(t)$ is the heat current and $\langle \mathbf{J}(0) \cdot \mathbf{J}(\tau)\rangle$ is the heat current auto-correlation function. For liquid system, the heat current has two contributions, $\mathbf{J}=\mathbf{J}^{\rm k}+\mathbf{J}^{\rm p}$.
The kinetic term is
$
\mathbf{J}^{\rm k} =\sum_{i}{\mathbf{v}_i E_i}
$,
where $E_i$ and $\mathbf{v}_i$ are the total energy and velocity of atom $i$, respectively. The potential term for many-body potentials is \cite{fan2021neuroevolution}
$
\mathbf{J}^{\rm p} = \sum_{i} \mathbf{W}_i \cdot \mathbf{v}_i
$,
where 
$
\mathbf{W}_i=\sum_{j \neq i}\mathbf{r}_{i j} \otimes \frac{\partial U_j}{\partial \mathbf{r}_{j i}}
$
is the virial tensor of atom $i$ and
$\mathbf{r}_{ij} = \mathbf{r}_j - \mathbf{r}_i$, $\mathbf{r}_i$ being the position of atom $i$.
According to the decomposition of the heat current, the thermal conductivity can be decomposed into three terms: 
$
\kappa(t) = \kappa^{\rm pp}(t) + \kappa^{\rm kk}(t) + \kappa^{\rm pk}(t)
$,
where the potential-potential term $\kappa^{\rm pp}$, the kinetic-kinetic term $\kappa^{\rm kk}$, and the cross term $\kappa^{\rm pk}$ correspond to the following heat current auto-correlation functions: $\langle \mathbf{J}^{\rm p}(0) \cdot \mathbf{J}^{\rm p}(\tau)\rangle$, $\langle \mathbf{J}^{\rm k}(0) \cdot \mathbf{J}^{\rm k}(\tau)\rangle$, and $\langle \mathbf{J}^{\rm p}(0) \cdot \mathbf{J}^{\rm k}(\tau)\rangle+\langle \mathbf{J}^{\rm k}(0) \cdot \mathbf{J}^{\rm p}(\tau)\rangle$. 

Besides, we use the homogeneous nonequilibrium \gls{md} method \cite{fan2019homogeneous} to calculate $\kappa^{\rm pp}$. In this method, an external driving force
$
\mathbf{F}_i^{\mathrm{ext}}= \mathbf{F}_{\mathrm{e}} \cdot \mathbf{W}_i
$
is exerted on each atom $i$, driving the system out of equilibrium. Here, $\mathbf{F}_{\mathrm{e}}$ (with magnitude $F_{\rm e}$) is the driving force parameter with the dimension of inverse length.
In this work, $F_{\mathrm{e}}$ was chosen as $0.001$ \AA$^{-1}$, which has been tested to be small enough to keep the system within the linear response regime.
The driving force will induce an ensemble-averaged steady-state non-equilibrium heat current $\mathbf{J}^{\rm p}$ (with magnitude $J^{\rm p}$) of the potential term, which is related to the thermal conductivity: $\kappa^{\rm pp} = \frac{J^{\rm p}}{TVF_{\rm e}}$.
The thermal conductivity can be further decomposed with respect to the vibrational frequency $\omega$ to obtain the spectral thermal conductivity $\kappa^{\rm pp}(\omega)$ \cite{fan2019homogeneous}.

\vspace{0.5cm}

\section*{Data availability}

The training and test datasets and the trained machine-learned potential models will be made available upon publication.

\section*{Code availability}

The source code for the graphics processing units molecular dynamics (GPUMD) package is available at \url{https://github.com/brucefan1983/GPUMD}.

\section*{Declaration of competing interest}
The authors declare that they have no competing interests.

\section*{Contributions}

KX trained the machine-learned models and performed the molecular dynamics simulations.
TL did the density functional theory calculations.
NX did the quantum chemistry level calculations.
PY did the training data sampling.
SC, NW, JX, and ZF supervised the project.
KX, LT, SC, and ZF drafted the manuscript.
All authors proofread the manuscript. 

\section*{Acknowledgments}
ZF was supported by the National Science and Technology Innovation 2030 Major Program from the Ministry of Science and Technology of China (No. 2024ZD0606900).
KX, TL, and JX acknowledge support from the National Key R\&D Project from the Ministry of Science and Technology of China (No. 2022YFA1203100), the Research Grants Council of Hong Kong (No. AoE/P-701/20), and RGC GRF (No. 14220022). 

\bibliography{refs}

\clearpage
\newpage

\newcommand{\stitlecaptionlabel}[4]{
    \phantomsection
    \addcontentsline{toc}{subsection}{\ref{#4}. #1}
    \caption[{#1}]{\textbf{#2.} #3}
    \label{#4}
}
\newcommand{\titlecaptionlabel}[3]{
    \stitlecaptionlabel{#1}{#1}{#2}{#3}
}

\newcommand{\manuallabel}[2]{\def\@currentlabel{#2}\label{#1}}
\makeatother

\newcounter{note}
\newcommand{\notetitlelabel}[2]{
    \phantomsection
    \refstepcounter{note}
    \addcontentsline{toc}{subsection}{\ref{#1}. #2}
    \manuallabel{#1}{\thenote}
    \subsection*{Supplemental Note \thenote: #2}
}
\newcommand*{\noteautorefname}{Supplementary Note}

\setcounter{secnumdepth}{0}
\renewcommand{\refname}{}
\renewcommand{\figurename}{Figure}
\renewcommand{\tablename}{Table}
\renewcommand\thefigure{S\arabic{figure}}
\renewcommand\thetable{S\arabic{table}}
\renewcommand\thenote{S\arabic{note}}
\renewcommand\theequation{S\arabic{equation}}
\renewcommand{\topfraction}{.9}
\renewcommand{\bottomfraction}{.9}
\renewcommand{\floatpagefraction}{.9}
\newcommand{\listreferencename}{Supplementary References}
\def\sectionautorefname{Sect.}
\def\figureautorefname{Fig.}
\def\tableautorefname{Table}
\def\equationautorefname{Eq.}

\renewcommand{\bibsection}{\section*{\listreferencename}}
\renewcommand{\vec}[1]{\ensuremath\boldsymbol{#1}}
\newcommand{\dd}{\mathrm{d}}
\providecommand{\todo}[1]{\textbf{\color{red}#1}}

\setcounter{figure}{0}
\setcounter{table}{0}

\onecolumngrid
\begin{center}
\huge\textbf{Supplementary Information:}
\end{center}



\begin{figure}[b]
\centering
\includegraphics[width=0.9\columnwidth]{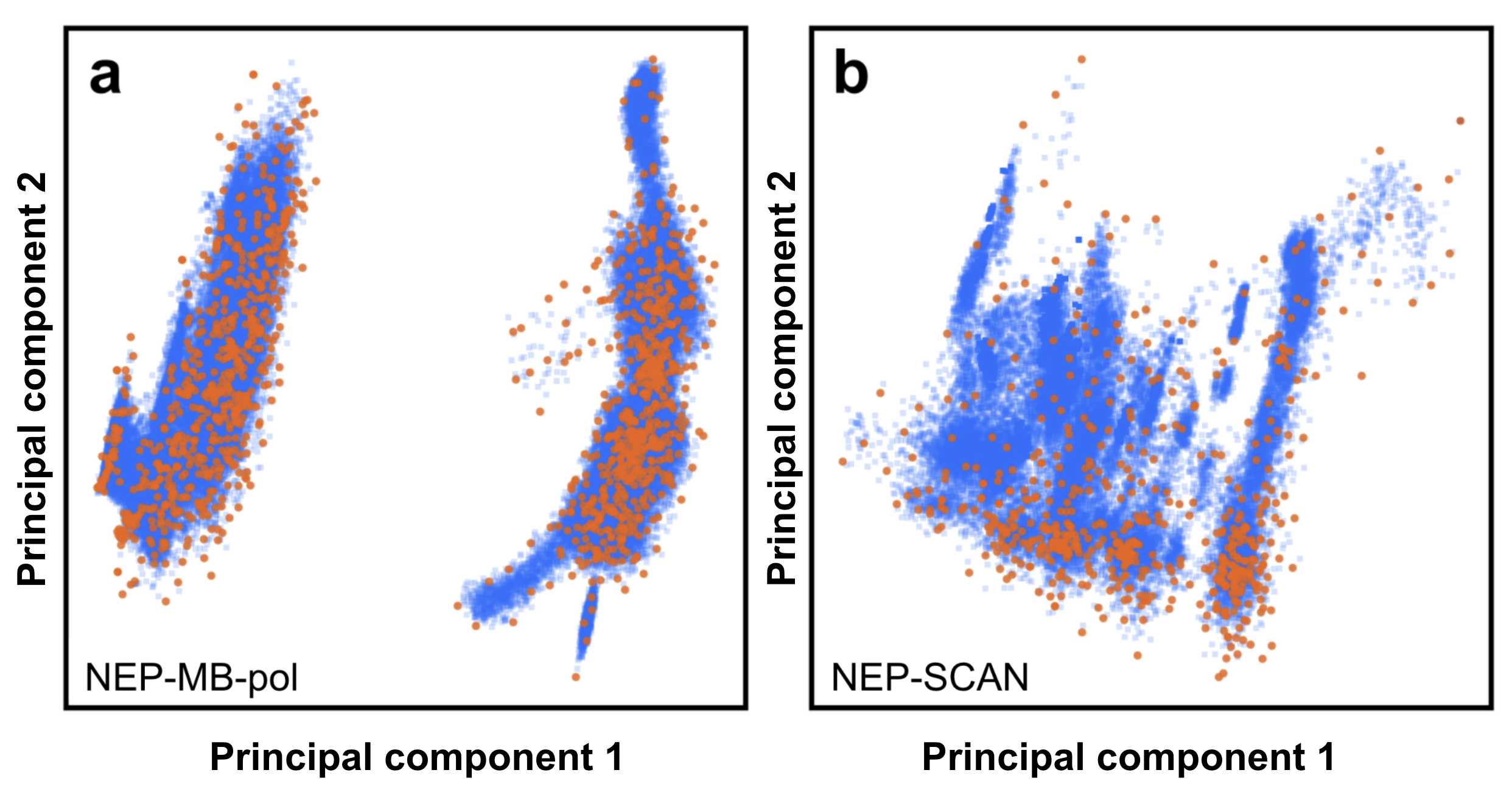}
\caption{
\textbf{Training data subsampling.}
Representative training datasets were selected from the original extensive datasets \cite{Zhai2023JCP,Zhang2021PRL} using farthest point sampling in descriptor space and visualized via principal component analysis for (a) the neuroevolution potential (NEP) trained on many-body polarization (MB-pol) reference dataset  and (b) the NEP trained on the strongly constrained and appropriately normed (SCAN) functional reference dataset. Blue points represent the original full datasets, while orange points indicate the selected representative datasets, comprising about 1\% of the original full datasets.}
\label{fig:S1-fps-pca}
\end{figure}

\begin{figure}[b]
\centering
\includegraphics[width=0.6\columnwidth]{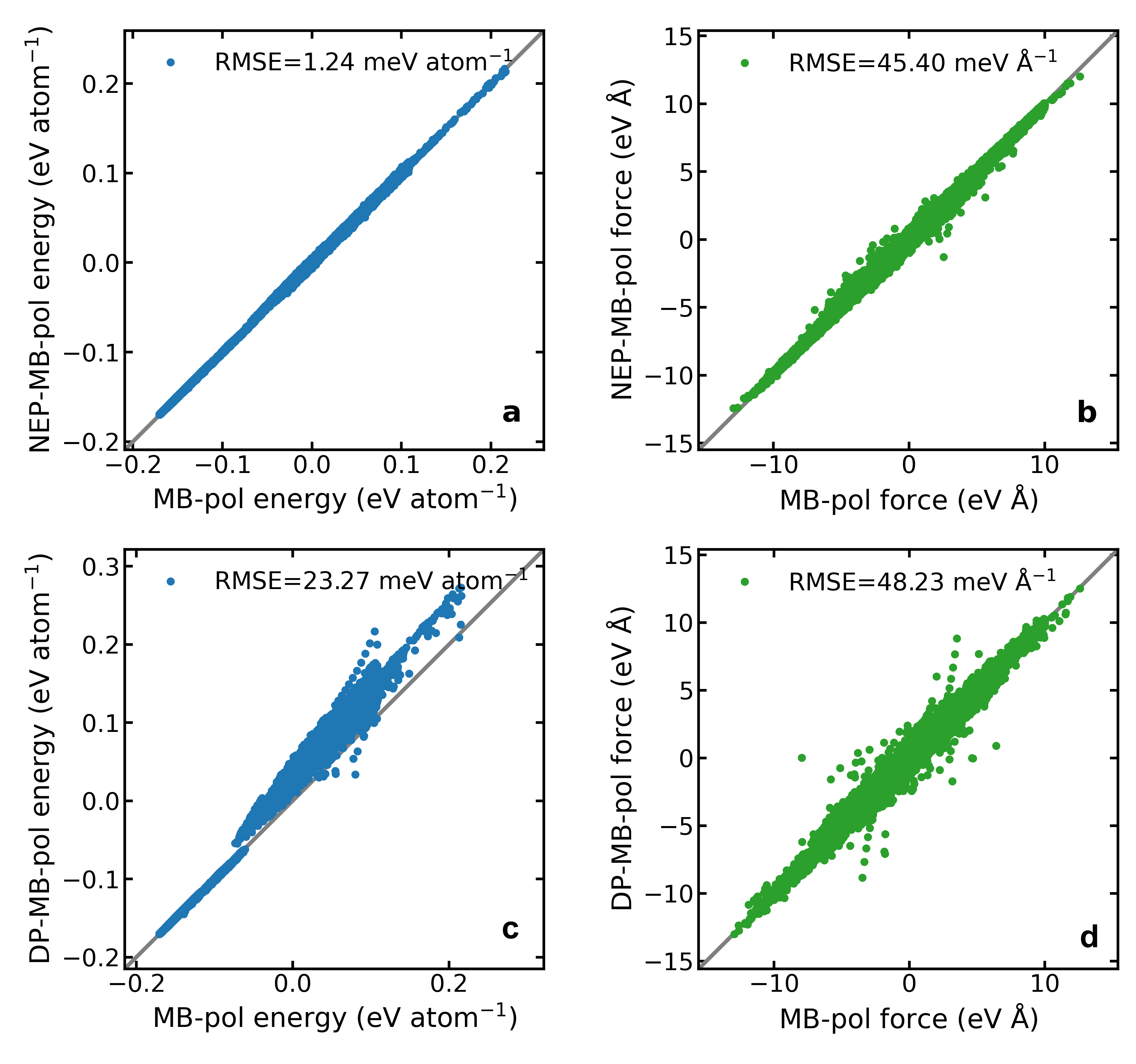}
\caption{
\textbf{Parity plots for MB-pol dataset.}
Parity plots comparing energy and force predictions from (a-b) neuroevolution potential (NEP) and (c-d) deep potential (DP) trained on many-body polarization (MB-pol) reference datasets \cite{Zhai2023JCP}.}
\label{fig:S2-nep-mb-pol_DP-mb-pol}
\end{figure}


\begin{figure}[b]
\centering
\includegraphics[width=\columnwidth]{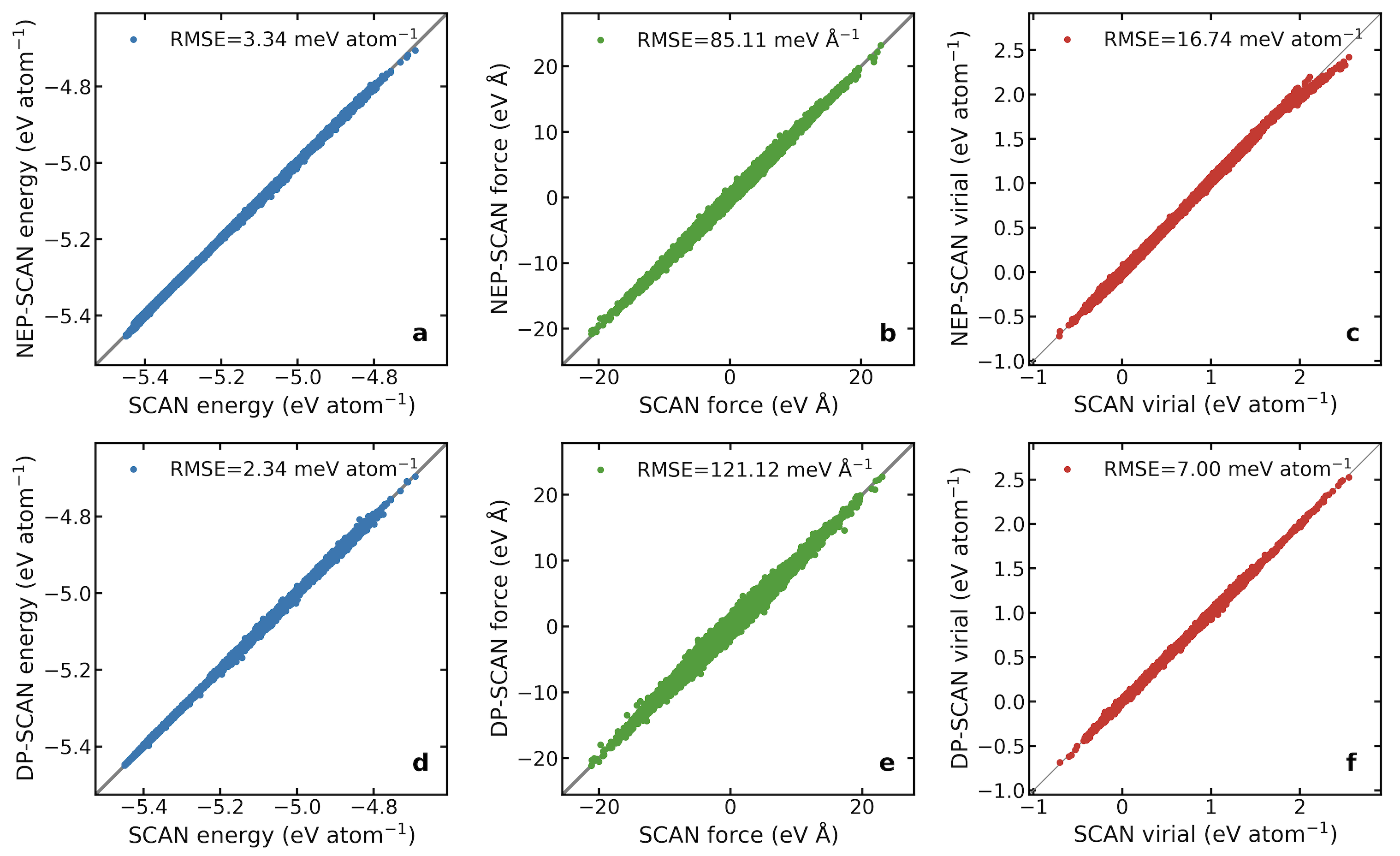}
\caption{
\textbf{Parity plots for SCAN dataset.}
Parity plots comparing energy, force, and virial predictions from (a-b) neuroevolution potential (NEP) and (c-d) deep potential (DP) trained on the strongly constrained and appropriately normed (SCAN) functional datasets \cite{Zhang2021PRL}.}
\label{fig:S3-nep-scan-V2-DP-SCAN}
\end{figure}

\begin{figure}[b]
\centering
\includegraphics[width=0.8\columnwidth]{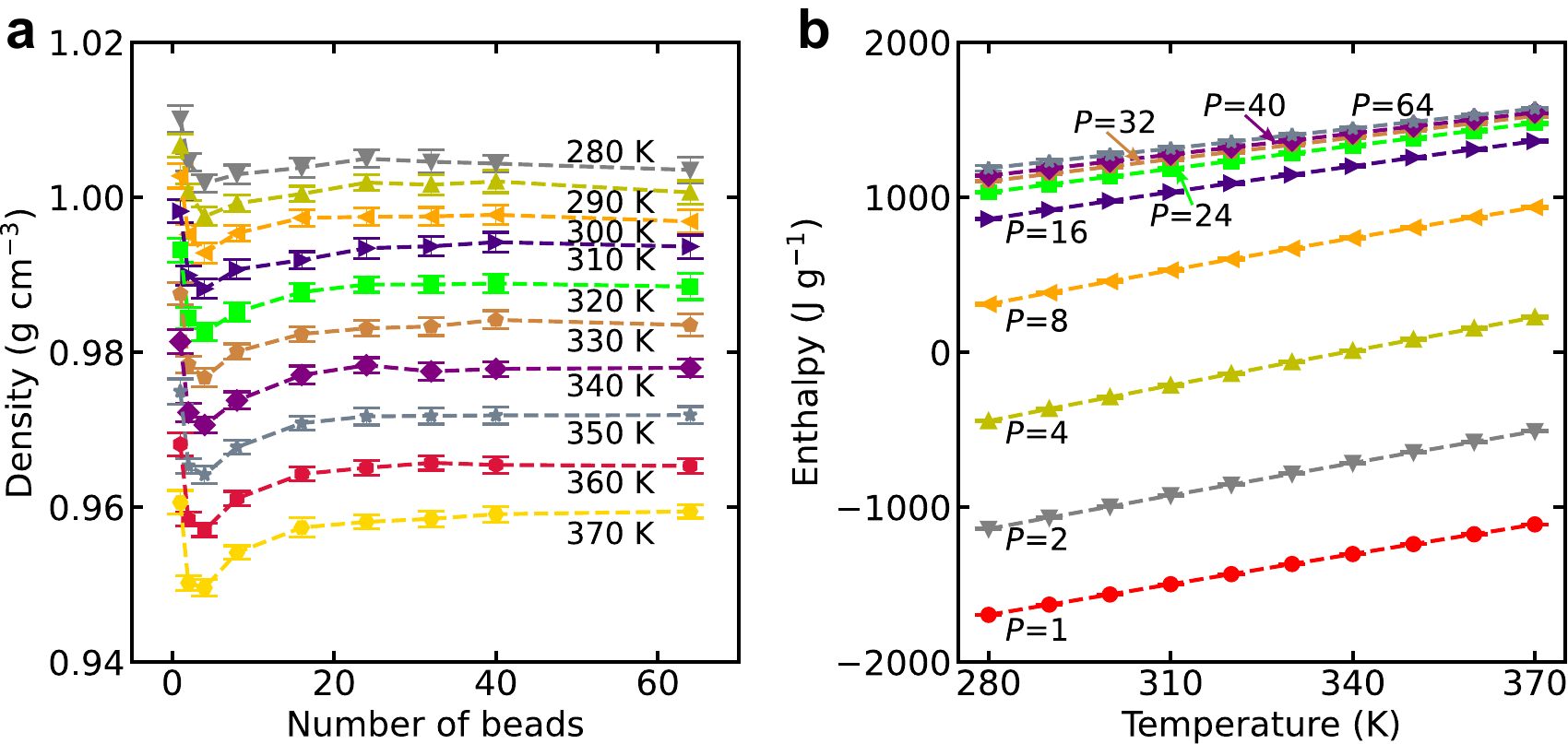}
\caption{
\textbf{Density and enthalpy of water.}
(a) Density of water as a function of the number of beads at various temperatures, calculated using path-integral molecular dynamics (PIMD) simulations with the neuroevolution potential trained on many-body polarization reference dataset (NEP-MB-pol). (b) System enthalpy as a function of temperature for different numbers of beads, $P$, obtained using PIMD simulations with the same potential. The pressure is kept at 1 atm.}
\label{fig:S4-dens}
\end{figure}

\begin{figure}[b]
\centering
\includegraphics[width=0.9\columnwidth]{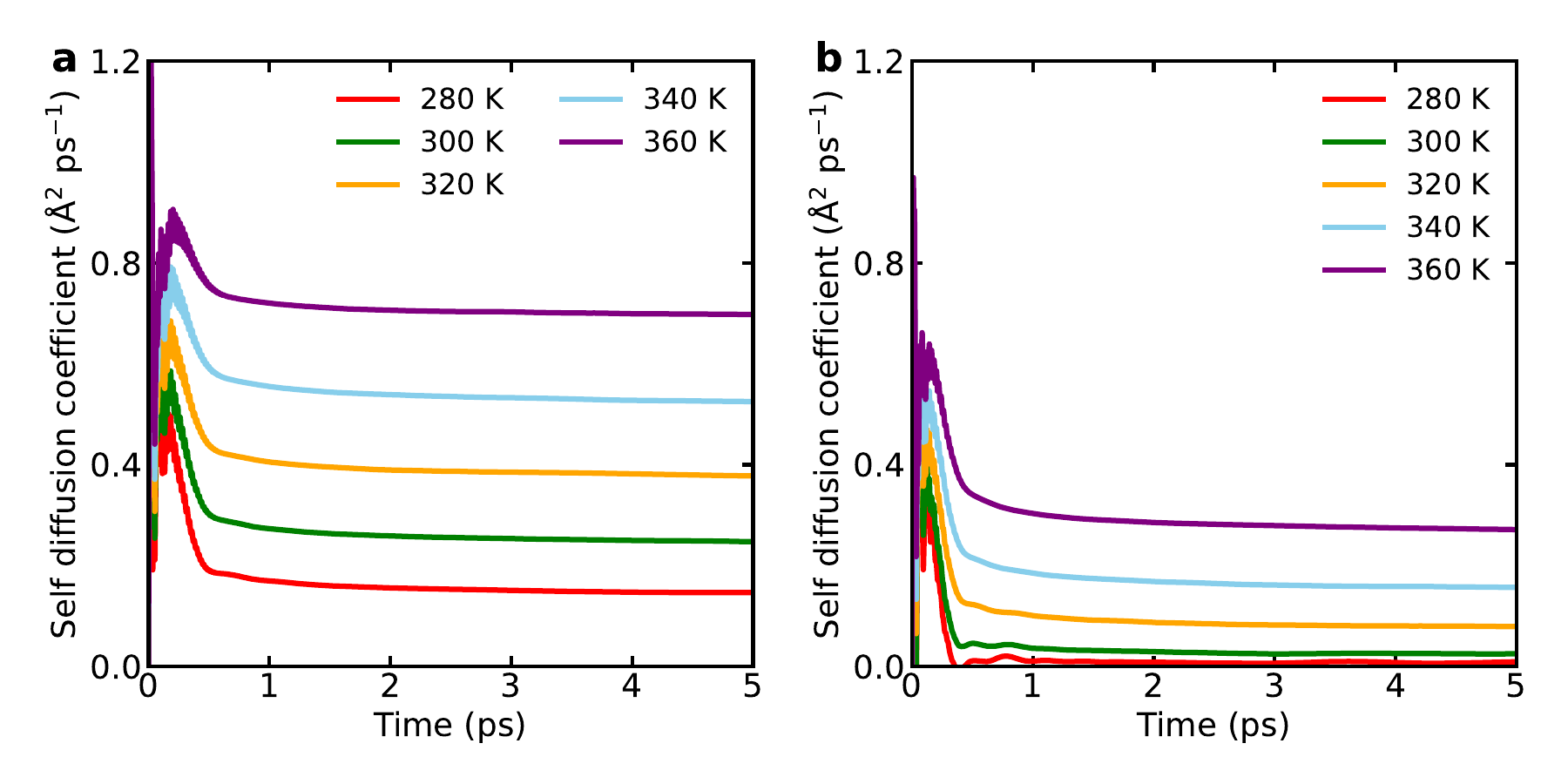}
\caption{
\textbf{Running self-diffusion coefficient of water.}
Convergence of the self-diffusion coefficient of water as a function of correlation time, calculated via thermostatted ring-polymer molecular dynamics simulations with 32 beads, using the neuroevolution potential trained on (a) the many-body polarization (NEP-MB-pol) and (b) the strongly constrained and appropriately normed functional (NEP-SCAN) reference datasets at various temperatures and 1 atm.}
\label{fig:S5-sdc}
\end{figure}

\begin{figure}[b]
\centering
\includegraphics[width=0.8\columnwidth]{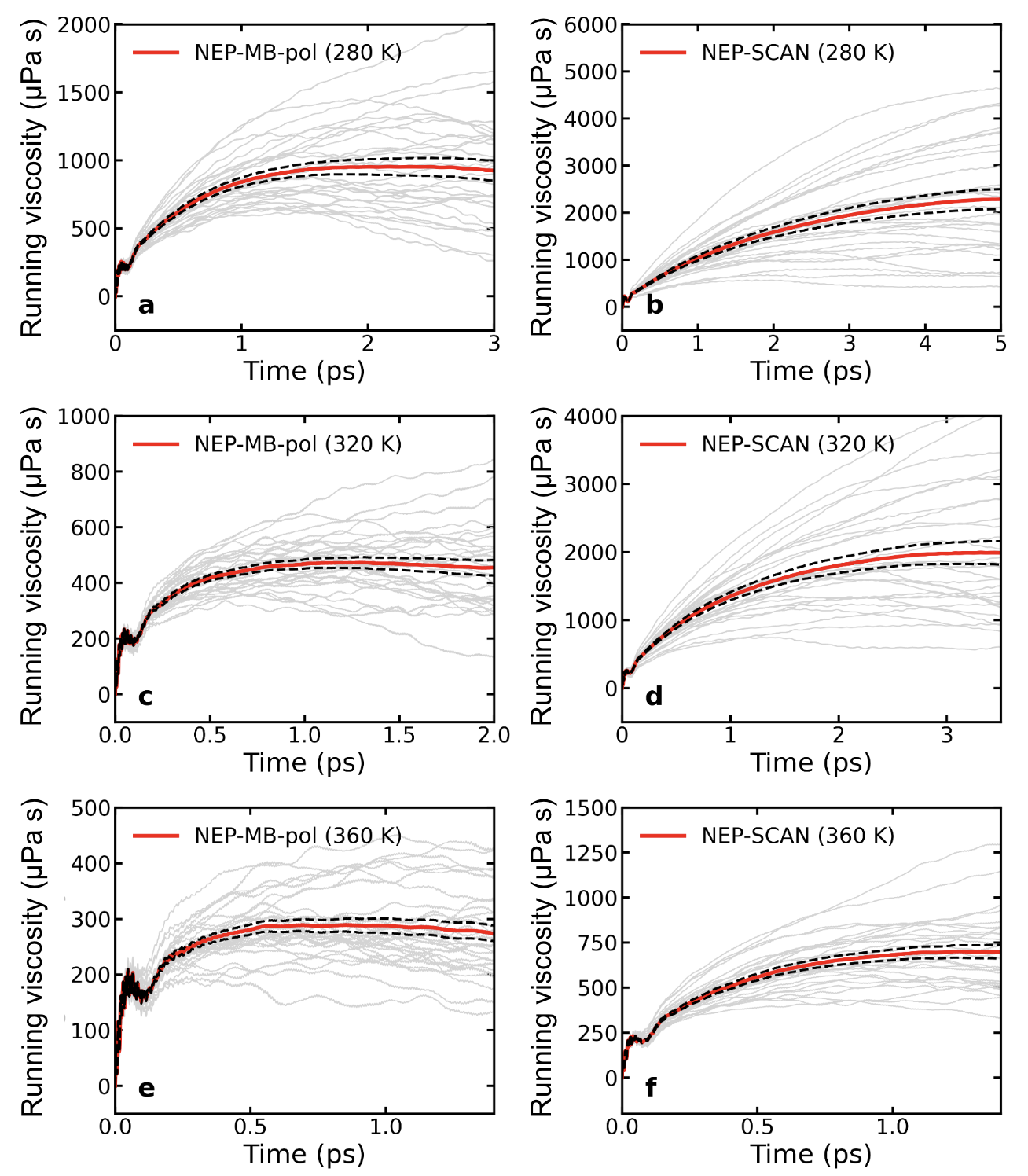}
\caption{
\textbf{Running viscosity of water.}
Convergence of viscosity of water as a function of correlation time, calculated via thermostatted ring-polymer molecular dynamics (TRPMD) simulations with 32 beads, using neuroevolution potential trained on (a,c,e) the many-body polarization (NEP-MB-pol) and (b,d,f) the strongly constrained and appropriately normed functional (NEP-SCAN) reference datasets at various temperatures and 1 atm.}
\label{fig:S6-visc}
\end{figure}

\begin{figure}[b]
\centering
\includegraphics[width=0.9\columnwidth]{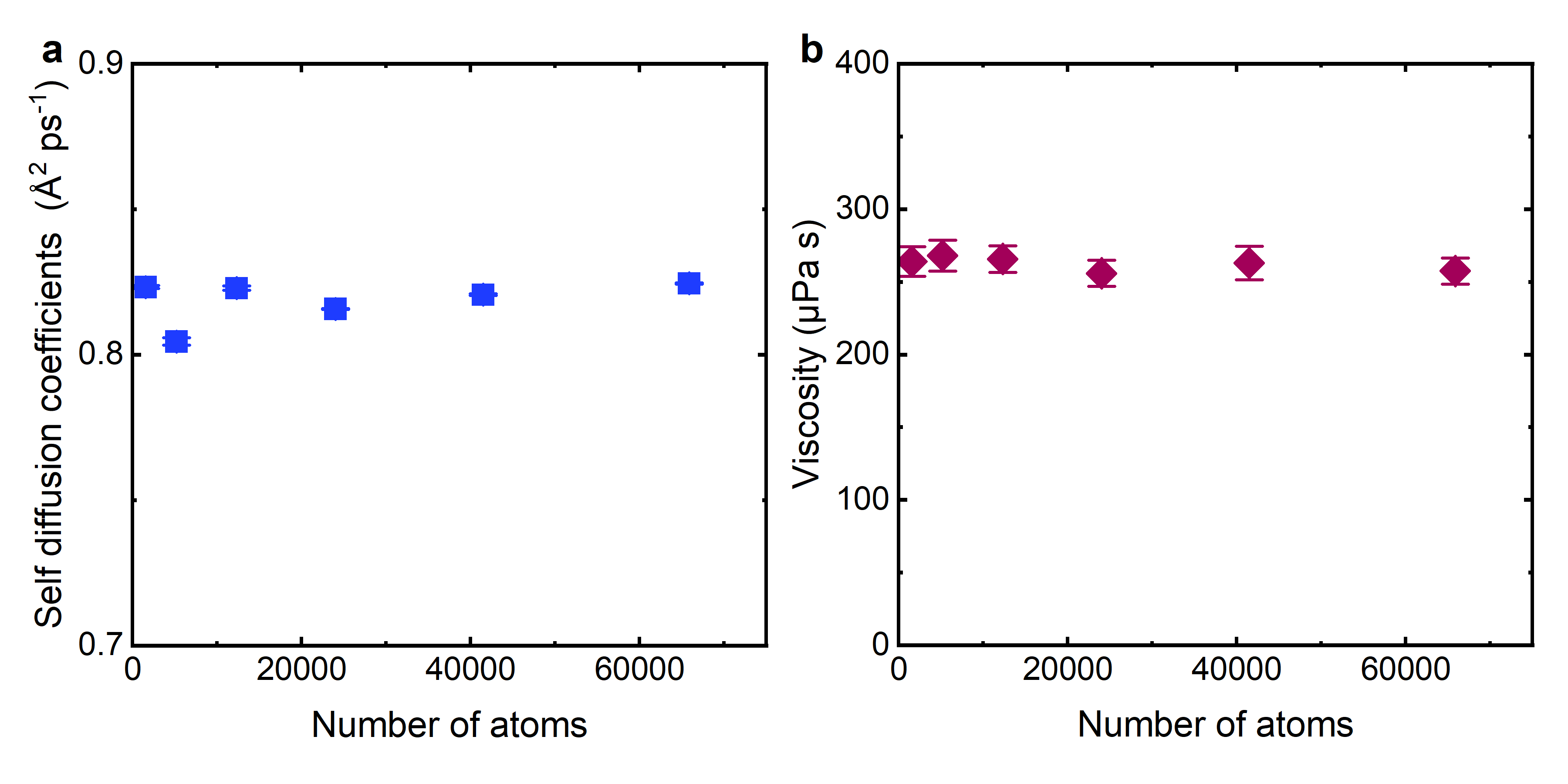}
\caption{
\textbf{Size convergence.}
Convergence of (a) self-diffusion coefficient and (b) viscosity of water as a function of system size (number of atoms), calculated via thermostatted ring-polymer molecular dynamics (TRPMD) simulations with 32 beads, using NEP-MB-pol at 300 K and 1 atm.}
\label{fig:S7-Size_effect}
\end{figure}

\begin{figure}[b]
\centering
\includegraphics[width=0.8\columnwidth]{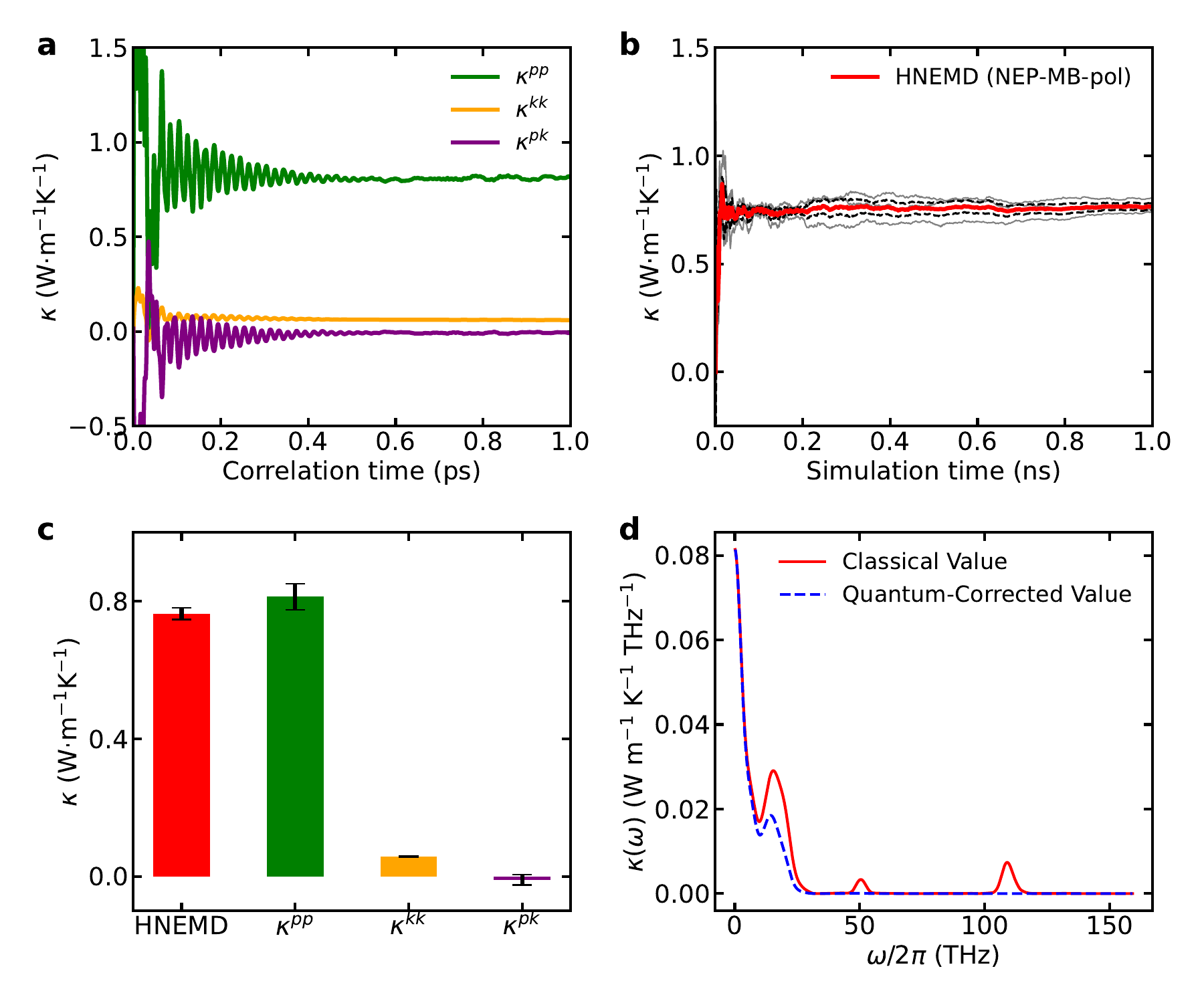}
\caption{
\textbf{Thermal conductivity.}
(a) Convergence of water's thermal conductivity as a function of correlation time t, calculated using equilibrium molecular dynamics (EMD) with NEP-MB-pol at 300 K and 1 atm, showing contributions from the potential-potential term ($\kappa^{pp}$), the kinetic-kinetic term ($\kappa^{kk}$), and the cross term ($\kappa^{pk}$), respectively. (b) Thermal conductivity (potential-potential term) of water as a function of simulation time obtained using homogeneous nonequilibrium molecular dynamics (HNEMD) method. (c) Comparison of thermal conductivity values from EMD and HNEMD methods. (d) Classical and quantum-corrected spectral thermal conductivity (potential-potential term) for water.}
\label{fig:S8-kappa}
\end{figure}

\end{document}